\documentclass[preprint,12pt,authoryear]{elsarticle}

\usepackage{amssymb}
\usepackage{amsmath}
\usepackage{multirow}
\usepackage{tabularx}
\usepackage{siunitx}
\usepackage{natbib}
\usepackage{xcolor}
\usepackage{subcaption}
\usepackage{url}

\myfooter[C]{Accepted manuscript, not the publisher's version of record. \copyright\ 2026, CC BY-NC-ND 4.0. Ocean Engineering 365 (2026) 127187.}
\myfooterfont{\scriptsize}
\makeatletter
\def\ps@elsaccepted{%
  \let\@oddhead\@empty
  \let\@evenhead\@empty
  \def\@oddfoot{\hbox to \textwidth{\hfil\@myfooterfont\@elsarticlemyfooter\hfil}}%
  \let\@evenfoot\@oddfoot
}
\g@addto@macro\ps@pprintTitle{\def\@oddfoot{\hbox to \textwidth{\hfil\@myfooterfont\@elsarticlemyfooter\hfil}}\let\@evenfoot\@oddfoot}
\makeatother
\AtBeginDocument{\pagestyle{elsaccepted}}

\journal{Ocean Engineering}

\begin{document}

\begin{frontmatter}

    \title{On the Limits of Univariate Deep Learning for Significant Wave Height Forecasting} 

    \author[1]{Yilin Zhai}
    \author[0,1,2]{Hongyuan Shi \corref{cor1}} 
    \author[1,3]{Zaijin You}
    \cortext[cor1]{Corresponding author}

    \affiliation[0]{
        organization={Shandong Key Laboratory Of Estuary And Coast \& Nuclear Environment},
        addressline={Ludong University},
        city={Shandong},
        postcode={264025},
        state={Yantai},
        country={China}
    }
    \affiliation[1]{
        organization={School of Hydraulic and Civil Engineering},
        addressline={Ludong University},
        city={Shandong},
        postcode={264025},
        state={Yantai},
        country={China}
    }

    \affiliation[2]{
        organization={Institute of Coastal Research},
        addressline={Ludong University},
        city={Yantai},
        postcode={264025},
        state={Shandong},
        country={China}
    }
    \affiliation[3]{
        organization={Navigation College},
        addressline={Dalian Maritime University},
        city={Dalian},
        postcode={116026},
        state={Liaoning},
        country={China}
    }

    \begin{abstract}
        This study conducts a systematic hyperparameter search across five deep learning architectures---DLinear, LSTM, PatchTST, ResAttLstm, and Mamba2---and nine context lengths (1--168~h) for single-station significant wave height ($H_s$) forecasting on NDBC buoy 41009, followed by re-evaluation of the best configurations on a 47-buoy, 37-year corpus. The five families converge to a common performance level on the multi-buoy evaluation (between-family SD = $0.0014$~m$^2$, 0.8\% of the grand mean), a spread dwarfed by the $4.83\times$ cross-dataset MSE shift between buoy corpora. All multi-buoy trials beat persistence (mean skill $+0.062$), but no architecture consistently outperforms the others. On the single-buoy experiment, skill peaks at 12--24~h where five trials fall below persistence, per-family Q4/Q3 test MSE ratios range from $2.4$ to $2.6$, and deep models underperform persistence for the most extreme 1\% of waves. These findings are consistent with the interpretation that persistence already captures the dominant linear-inertial signal in univariate $H_s$, and that architecture engineering under this univariate input setting has reached diminishing returns: cross-buoy variance, not model class, dominates forecast error. Future work should prioritise atmospheric covariates, zero-shot cross-buoy transfer, and decomposition of $H_s$ into swell and wind-sea components. By establishing a rigorous reference baseline for what univariate $H_s$ models can and cannot achieve, this study provides a benchmark against which future multivariate and physics-informed approaches can be calibrated, and offers practical guidance for lightweight buoy-level forecasting in mid-latitude storm-dominated and swell-mixed environments.

    \end{abstract}


    \begin{keyword}
        Significant wave height \sep Deep learning \sep Time series forecasting \sep Univariate model \sep NDBC buoy \sep Model comparison



    \end{keyword}

\end{frontmatter}

\section{Introduction}
\label{sec:intro}
Significant wave height ($H_s$) is the primary variable for ship routing, offshore operations, coastal flood warning, and harbour safety. Wave transformation in harbour and nearshore environments has been extensively characterised through phase-resolving numerical models \citep{gao2021investigation,gao2023mechanism}. Operational forecasting relies on numerical wave models (WAVEWATCH III \citep{tolman2009user}; SWAN \citep{booij1999third}) forced by atmospheric fields, achieving correlation coefficients of $0.95$--$0.96$ at Day~1 \citep{campos2024development}; however, these systems require global atmospheric forcing and substantial computational resources, limiting their deployment for lightweight, buoy-level products. This gap has motivated a parallel line of work: models that forecast $H_s$ solely from a buoy's own wave record, without atmospheric covariates.

Since 2018, a series of studies has applied pure time-series deep learning to the SWH prediction task---models whose sole input is a history of $H_s$ values and whose output is a future $H_s$ value. Before the deep learning era, traditional statistical time series methods---including ARIMA, VAR, and GARCH-type models---were also applied to $H_s$ forecasting; these linear approaches provide well-calibrated forecasts with minimal data requirements but cannot capture the non-linear wave-field evolution that has motivated the architectures surveyed below. Recurrent architectures, led by LSTM \citep{hochreiter1997long}, were the first to be evaluated on NDBC buoy data and demonstrated that learned temporal dependencies can modestly improve upon persistence. Transformer-based models designed for long-sequence forecasting---Informer \citep{zhou2021informer}, Autoformer \citep{wu2021autoformer}, FEDformer \citep{zhou2022fedformer}, PatchTST \citep{nie2022time}, and iTransformer \citep{liu2024itransformer}---have since been applied to SWH under the hypothesis that self-attention \citep{vaswani2017attention} can extract non-linear wave-field evolution over extended context windows. Multi-scale convolutional architectures such as TimesNet \citep{wu2022timesnet} and wavelet graph neural networks \citep{chen2021significant} capture periodicity at multiple temporal resolutions. Most recently, the Mamba family of selective state-space models \citep{gu2023mamba,dao2024transformers} has been introduced for SWH, alongside Chronos-based foundation models \citep{zhai2025improving}, time-frequency networks \citep{zhang2025novel}, Fourier neural operators \citep{hazarika2025swr}, and physics-informed reinforcement learning for wave routing \citep{bora2026physics}.

Jiang et al.\ \citep{jiang2024comment} recently challenged this premise, demonstrating that several complex architectures do not outperform a simple autoregressive model for univariate SWH. Their analysis, however, was restricted to two context lengths (6 and 24~h) on a small buoy set, leaving open whether the null result generalises to the long-context regime (24--168~h), where models must capture dependencies spanning tens to hundreds of time steps---a setting for which architectures such as PatchTST and Mamba2 were specifically designed.

Three gaps limit prior work. First, the long-context hypothesis has only been examined at 6~h and 24~h; no study has tested whether extending the input window to 96--168~h reveals additional non-linear SWH structure. Second, no controlled comparison pairs architectures released after 2016 alongside linear baselines under a unified hyperparameter optimisation protocol. Third, existing multi-architecture evaluations span at most 16~buoys over 2~years, leaving unclear whether single-buoy findings transfer to heterogeneous wave climates. To ensure realistic forecasting conditions, all evaluations in the present study employ strict chronological train-validation-test splits (70/15/15 by time); the baseline for all experiments is the persistence forecast (Section~\ref{sec:models}).

To determine whether the reported gains reflect genuine non-linear skill extraction or merely overfitting to local wave climates, this study addresses three research questions. (1)~Do modern architectures with substantially different inductive biases---DLinear \citep{Zeng2022AreTE} (linear decomposition), LSTM (recurrent), PatchTST (patched attention), ResAttLstm (CNN-attention-LSTM hybrid), and Mamba2 (state-space)---yield a meaningfully lower test MSE than persistence for univariate SWH forecasting, or do they converge to a common performance ceiling? (2)~Does extending the context window from 1~h to 168~h surface non-linear SWH structure that is invisible at short context lengths, or does performance saturate once the linear-inertial signal is captured? (3)~Do the single-buoy findings replicate when models are re-trained and evaluated on a heterogeneous 47-buoy, 37-year NDBC corpus? The study provides a systematic comparison of five forecasting paradigms under a unified hyperparameter search protocol, evaluates context-length scaling from 1~h to 168~h across architectures, and reports a large-scale evaluation on 47~NDBC buoys spanning 37~years (1988--2025) with strict chronological splits---an order of magnitude larger than prior controlled comparisons.

Answering these questions would establish whether the dominant research strategy in the deep-learning-for-SWH literature---architecture engineering---yields meaningful gains over persistence, or whether further model innovation has reached diminishing returns. A null result would imply that the principal bottleneck is not model class but the information content of univariate $H_s$ alone, redirecting future effort toward atmospheric data fusion and cross-buoy generalisation.

Section~\ref{method} describes the experimental design, datasets, task formulation, models, and evaluation protocol. Section~\ref{results} reports the results, and Sections~\ref{discussion} and~\ref{sec:conclusion} discuss the findings and conclude.

\section{Methodology}
\label{method}
\subsection{Experimental design}
\label{exp_design}
\begin{figure}[ht]
    \centering
    \includegraphics[width=\textwidth]{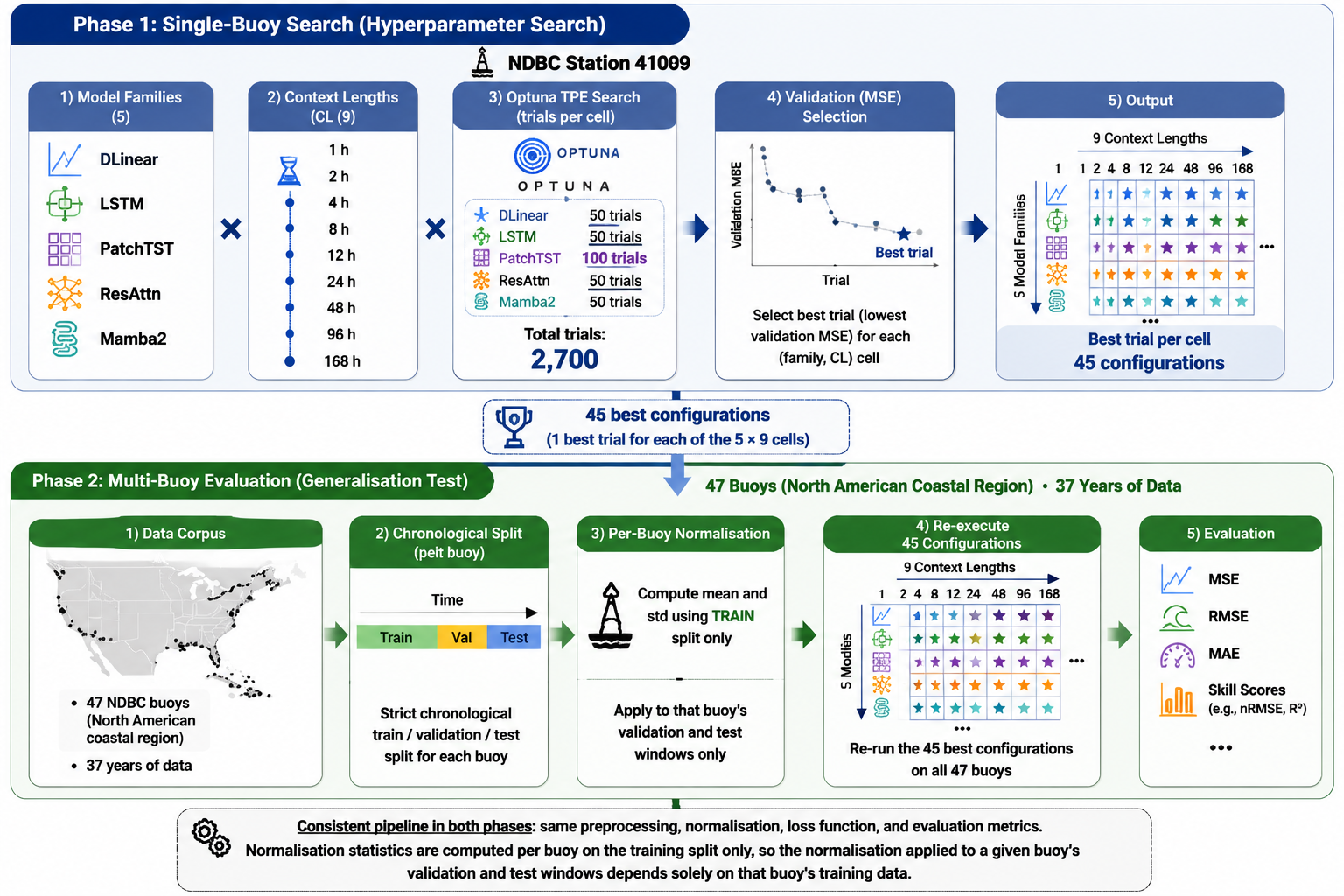}
    \caption{Overview of the two-phase experimental design. This figure is a conceptual schematic generated with GPT Image~2.0 and contains no scientific data.}
    \label{fig:exp_design}
\end{figure}

The study follows a two-phase design (Figure~\ref{fig:exp_design}): Phase 1 (single-buoy search): a controlled hyperparameter search is conducted on NDBC station 41009 alone, sweeping 5 model families × 9 context lengths (CL). DLinear, LSTM, PatchTST, and Mamba2 received 50 Optuna TPE trials per (family, $\text{CL}$) cell; ResAttLstm, owing to its larger hyperparameter space, received 100 trials per cell. The total is 2,700 trials across the five-family sweep. The single station eliminates cross-buoy variance, enabling a clean comparison of architectures and context lengths under identical data conditions. The best trial per (family, $\text{CL}$) cell—45 in total—is selected by validation MSE.

Phase 2 (multi-buoy evaluation): the 45 best configurations are re-executed on a 47-buoy corpus spanning three ocean basins and 37 years, with strict chronological train/validation/test splits. This phase tests whether the single-buoy findings generalise to a heterogeneous, at-scale deployment. The same preprocessing pipeline, normalisation, loss function, and evaluation metrics are used in both phases. Normalisation statistics (median and IQR) were computed on the combined multi-buoy training set after concatenation of all buoys' training splits; no cross-buoy information leakage occurred because normalisation parameters were derived exclusively from the training partition.

\subsection{Datasets}
\label{dataset}
\begin{figure}[ht]
    \centering
    \includegraphics[width=0.8\textwidth]{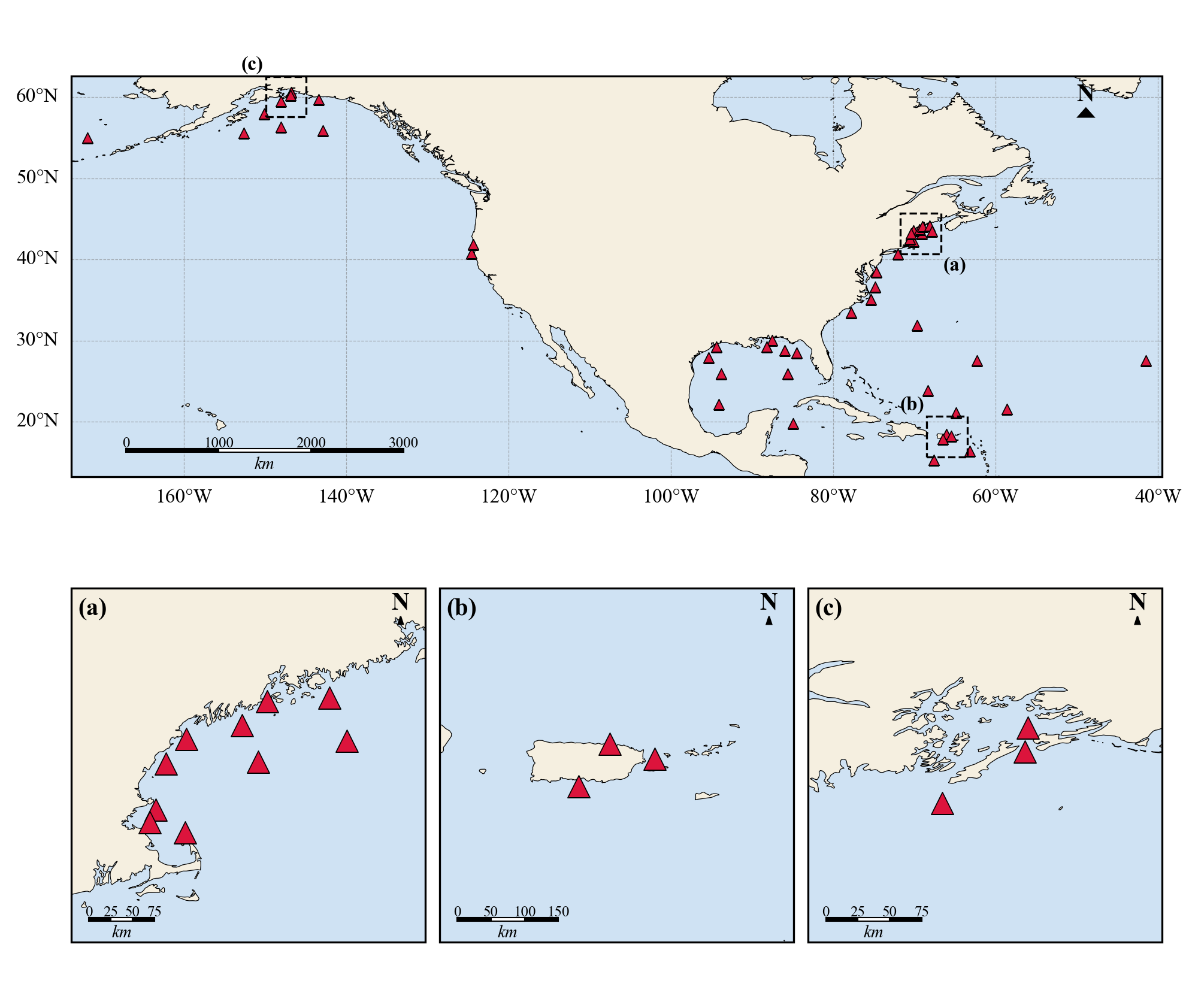}
    \caption{Geographic distribution of the 47 NDBC buoys used in the multi-buoy evaluation.}
    \label{fig:buoys}
\end{figure}

The single-buoy dataset is NDBC station 41009 (NDBC, 2025) (Atlantic Ocean, off Florida east coast, 28.5°N, 80.2°W, 30 m depth), spanning 1988-2025 at hourly resolution. The multi-buoy dataset is a corpus of 47 NDBC buoys spanning the mid-Atlantic, Pacific, and Gulf of Alaska, 1988-2025, hourly resolution. The buoy set (Figure~\ref{fig:buoys}) comprises 22 Atlantic buoys (from Florida to Nova Scotia), 18 Pacific buoys (California to the Aleutians), and 7 Gulf of Alaska buoys. Record lengths range from 12 to 37 years (median 33 years), yielding approximately 1.2 × $10^7$ total hourly samples. The same preprocessing pipeline is applied per buoy. Key dataset statistics are summarised in Table~\ref{tab:dataset_summary}.

\begin{table}[ht]
    \scriptsize
    \centering
    \begin{tabular}{lccccccccc}
        \hline
        Property & mean $\pm$ SD     & median & IQR    & Q1--Q3           & p95    & p99    & p99.9  & max     \\
        \hline
        41009    & $1.18 \pm 0.67$ m & 1.01 m & 0.79 m & $0.70$--$1.49$ m & 2.45 m & 3.48 m & 5.19 m & 9.79 m  \\
        47 buoys & $1.46 \pm 1.01$ m & 1.21 m & 1.08 m & $0.78$--$1.86$ m & 3.44 m & 5.10 m & 7.35 m & 16.91 m \\
        \hline
    \end{tabular}
    \caption{Dataset summary. All SWH statistics are computed directly from raw NDBC hourly records.}
    \label{tab:dataset_summary}
\end{table}

\subsection{Task and data preparation}
\label{sec:preprocessing}
The task is multi-step regression of SWH at lead times of 1-6 h from a context window of past $H_s$ values:

\begin{equation}
    [H_s(t+1), \ldots, H_s(t+6)] = f\bigl(H_s(t), H_s(t-1), \ldots, H_s(t-\text{CL}+1)\bigr)
\end{equation}

where $\text{CL} \in \{1, 3, 6, 12, 24, 48, 72, 96, 168\}$ hours.All six output steps are predicted simultaneously; the reported metrics are averaged over the six lead times unless otherwise noted. No exogenous covariates are provided; the sole input is $H_s$. All five families are evaluated under the same set of context lengths. $\text{CL} = 1$ provides the minimal-history setting, where models have access only to the most recent observation.

The time series was split into contiguous segments at gaps where the inter-sample interval exceeded one hour; segments shorter than $\text{CL} + 6$ were discarded. Zero SWH values were retained. The overall missing-data rate was below 2\% for station 41009 and below 5\% across the 47-buoy corpus. All preprocessing was applied per-buoy and prior to train-validation-test splitting.

\begin{figure}[ht]
    \centering
    \includegraphics[width=0.8\textwidth]{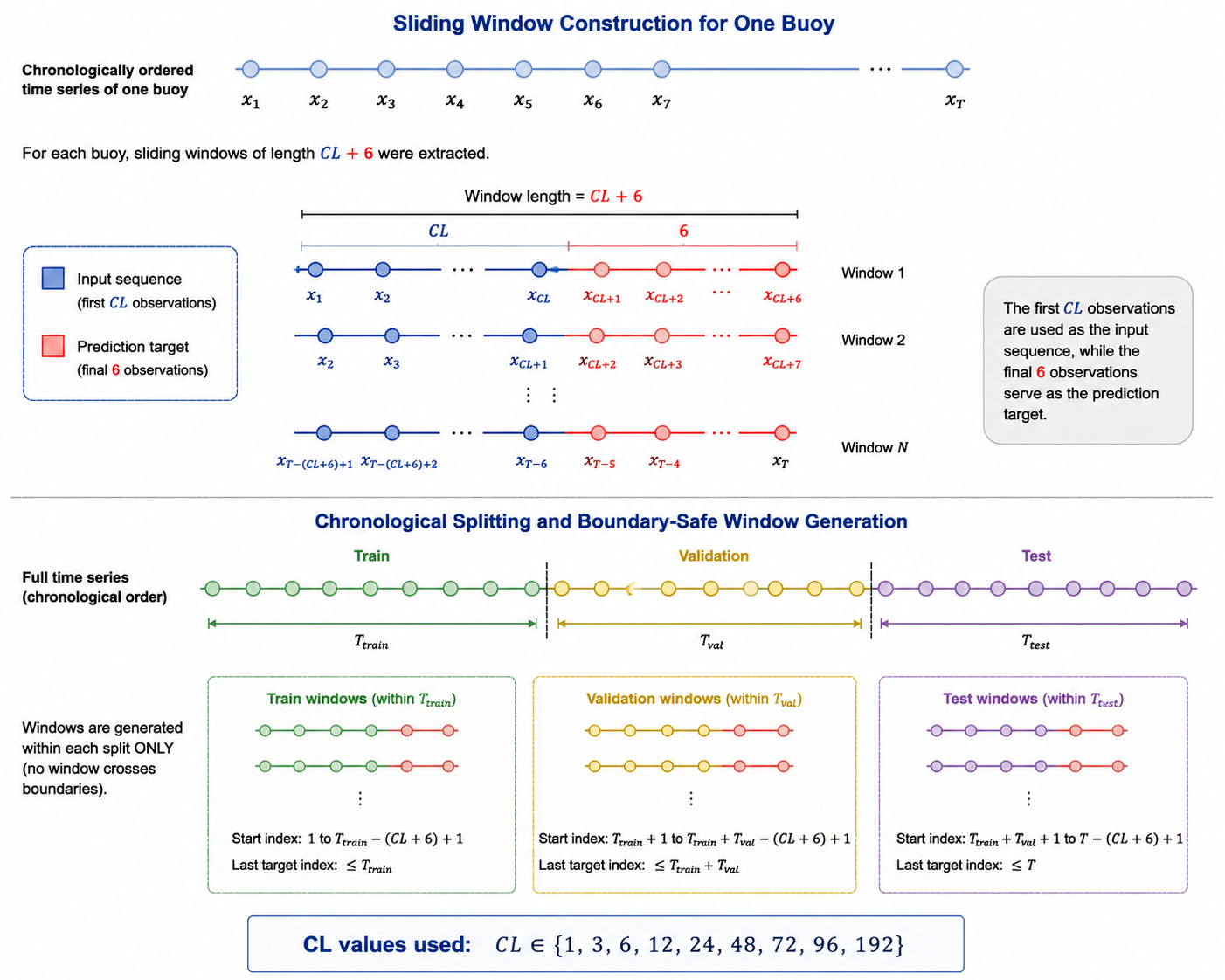}
    \caption{Schematic of the sliding-window construction. This figure is a conceptual schematic generated with GPT Image~2.0 and contains no scientific data.}
    \label{fig:windowing}
\end{figure}

All data are split chronologically 70/15/15 (train/validation/test). For each buoy, sliding windows of length CL + 6 are extracted: the first CL steps form the input, the final six steps are the targets (Figure~\ref{fig:windowing}). Windows are generated only after chronological partitioning; consequently, no window spans a train-validation or validation-test boundary.

Inputs and targets are normalised via:

\begin{equation}
    x' = \ln(1 + x), \qquad
    x'' = \frac{x' - \operatorname{median}(x')}{\operatorname{IQR}(x')}
\end{equation}

where median and IQR are computed on the training set. The log-transform reduces the positive skew of the SWH distribution; the +1 offset accommodates near-zero observations. Robust scaling (median/IQR) was preferred to z-score normalisation given the heavy-tailed distribution of SWH and the presence of infrequent extreme events that would otherwise distort the training mean and variance. The loss function is MSE on the transformed, scaled values, averaged over the six output steps. MSE, $R^2$, and Pearson correlation are also reported as test-set evaluation metrics:

\begin{gather}
    \text{MSE} = \frac{1}{n}\sum_{i=1}^{n}(y_i - \hat{y}_i)^2, \\
    R^2 = 1 - \frac{\sum_{i=1}^{n}(y_i - \hat{y}_i)^2}{\sum_{i=1}^{n}(y_i - \bar{y})^2}, \\
    r = \frac{\sum_{i=1}^{n}(y_i - \bar{y})(\hat{y}_i - \bar{\hat{y}})}{\sqrt{\sum_{i=1}^{n}(y_i - \bar{y})^2 \sum_{i=1}^{n}(\hat{y}_i - \bar{\hat{y}})^2}}
\end{gather}

where $y_i$ and $\hat{y}_i$ denote the true and predicted values in normalised space, and $\bar{y}$, $\bar{\hat{y}}$ are their respective means. The loss is computed in the normalised space; for reporting, all metrics (MSE, $R^2$, $r$, and persistence skill) are computed after inverse-transforming predictions back to the original SWH units (metres) via the inverse of the log--median/IQR pipeline, so that all reported MSE values are in m\textsuperscript{2} and directly interpretable against operational benchmarks. The median and IQR used for inverse transformation are those of the training set.

\subsection{Models, training, and evaluation}
\label{sec:models}
Five families spanning distinct inductive biases are evaluated.

\textbf{DLinear} \citep{Zeng2022AreTE}. Decomposition-linear: a moving-average trend branch and a seasonal branch, each a single linear projection.

\textbf{LSTM} \citep{hochreiter1997long}. Stacked LSTM with optional bidirectionality in intermediate layers; LSTM-based models have been validated for spectral wave evolution at coastal field sites \citep{wang2023field}. The final prediction layer remained unidirectional in all selected configurations.

\textbf{PatchTST} \citep{nie2022time}. Channel-independent patch transformer with learnable patch embedding. At context lengths $\text{CL} \le 16$, shorter than the default patch length, patchify is bypassed and a linear projection maps the input directly to the model dimension.

\textbf{ResAttLstm}. A hybrid architecture combining a residual CNN encoder with multi-head attention pooling and a unidirectional LSTM regression head, following the design pattern of \citep{mao2026attention}.

\textbf{Mamba2}~\citep{dao2024transformers}. Sequential stack of Mamba2 blocks with pre-norm and residual connections. At context lengths $\mathrm{CL} \leq 15$, a linear layer pads the sequence to 16 time steps to satisfy the SSD kernel's minimum sequence-length requirement.

The primary baseline is the persistence forecast: $\hat{y}(t+h) = y(t)$, for $h = 1, \ldots, 6$.

The single-buoy experiment employed Optuna TPE \citep{bergstra2011algorithms,akiba2019optuna} (seed = 42) with early stopping (patience 20, max 200 epochs), minimising validation MSE. The median pruner was configured with $n_{\text{startup}}=5$, $n_{\text{warmup\_steps}}=0$, and interval 1. All models used Adam \citep{kingma2014adam} ($\beta_1=0.9$, $\beta_2=0.999$, no weight decay) with cosine-annealing \citep{loshchilov2016sgdr} without restarts. Search spaces (Table~\ref{tab:hyperparam_search}) are identical across CL; the per-family trial budget is described in Section~\ref{exp_design}.

\begin{table}[ht]
    \centering
    \small
    \begin{tabularx}{\textwidth}{lcX}
        \hline
        Family     & \ Params & Key hyperparameters                                                                                                                                            \\
        \hline
        DLinear    & 2        & kernel\_size [9, 93]; lr [1e-3, 9.9e-3]                                                                                                                        \\
        LSTM       & 5        & n\_layers \{2,3\}; hidden \{64--256\}; bidirectional per layer; dropout [0.003, 0.49]; lr [1e-4, 1e-2]                                                         \\
        PatchTST   & 8        & patch\_len \{8--24\}; stride \{2--12\}; d\_model \{64--256\}; n\_layers \{1--6\}; n\_heads \{8,16\}; d\_ff \{128--512\}; dropout [0, 0.2]; lr [1e-5, 1e-2] log \\
        ResAttLstm & 7        & n\_cnn \{1--5\}; cnn\_hidden \{64--384\}; kernel \{3--7\}; attn\_heads \{4,8,16\}; lstm\_hidden \{64--256\}; bidirectional; lr [1e-5, 1e-2] log                \\
        Mamba2     & 6        & n\_blocks \{1--4\}; d\_model \{64--256\}; d\_state \{16--128\}; d\_conv \{2--4\}; expand \{1,2,4\}; lr [1e-5, 1e-2] log                                        \\
        \hline
    \end{tabularx}
    \caption{Continuous parameters are sampled uniformly (or log-uniform for learning rate). Categorical parameters are enumerated. Batch size was fixed at 4096 for all families.}
    \label{tab:hyperparam_search}
\end{table}

The 45 best configurations were re-trained on the combined multi-buoy training set and evaluated on the combined test set. All experiments were implemented in Python 3.13.9 with PyTorch 2.10. Hyperparameter search was managed with Optuna 4.8.0. All experiments (single-buoy and multi-buoy) were run on five A800 (PCIe) GPUs. Random seeds were fixed at 42 for Python, NumPy, and PyTorch, and CUDA deterministic mode was enabled. all NDBC buoy data are publicly accessible from the National Data Buoy Center (\url{https://www.ndbc.noaa.gov}).

\subsection{Evaluation metrics and statistical analysis}
The primary evaluation metric is persistence skill:
\begin{equation}
    \text{skill} = 1 - \frac{\operatorname{MSE}(\text{model})}{\operatorname{MSE}(\text{persistence})}
\end{equation}

where skill > 0 indicates improvement over the no-change baseline. Tail performance is assessed via $p_{99}$ persistence skill, computed on the 1\% of test samples with the largest SWH values.

All reported means are accompanied by standard errors ($\text{SE} = \text{SD} / \sqrt{n}$). Between-family standard deviation and the cross-dataset MSE ratio are reported as measures of effect size. A bootstrap 95\% confidence interval (10,000 resamples) is reported for the cross-dataset MSE ratio ($n = 45$ pairs) in Section~\ref{cross_dataset_transfer}. Formal null-hypothesis significance tests (e.g., Friedman, Nemenyi, Wilcoxon; \citep{demsar2006statistical}) were not applied because, with test sets exceeding $10^6$ samples, even trivially small architecture-driven differences would achieve statistical significance. Effect sizes directly quantify the practical magnitude of these differences, which is the question of interest here. All reported MSE values are in m\textsuperscript{2} (see Section~\ref{sec:preprocessing} for the inverse-transformation procedure). The $p_{99}$ tail metric is computed on the largest 1\% of test samples; for the single-buoy experiment this corresponds to approximately $1.1 \times 10^4$ samples per trial, and for the multi-buoy evaluation approximately $5.2 \times 10^5$ samples per trial, ensuring reasonable stability of the tail estimate.

\begin{equation}
    \sigma_{\text{between}} = \sqrt{\frac{1}{4}\sum_{i=1}^{5}\bigl(\bar{\varepsilon}_i - \bar{\varepsilon}\bigr)^2}, \qquad
    r = \frac{\bar{\varepsilon}_{\text{multi}}}{\bar{\varepsilon}_{\text{single}}}
\end{equation}

where $\bar{\varepsilon}_i$ denotes the per-family mean test MSE (averaged over nine context lengths), $\bar{\varepsilon}$ is the 5-family grand mean, and $\bar{\varepsilon}_{\text{multi}}$, $\bar{\varepsilon}_{\text{single}}$ are the grand mean test MSE values on the multi-buoy and single-buoy evaluations, respectively.

\clearpage
\section{Results}
\label{results}
\subsection{Hyperparameter Search}
\label{sec:hp_search}

\begin{figure}[ht]
    \centering
    \includegraphics[width=0.8\textwidth]{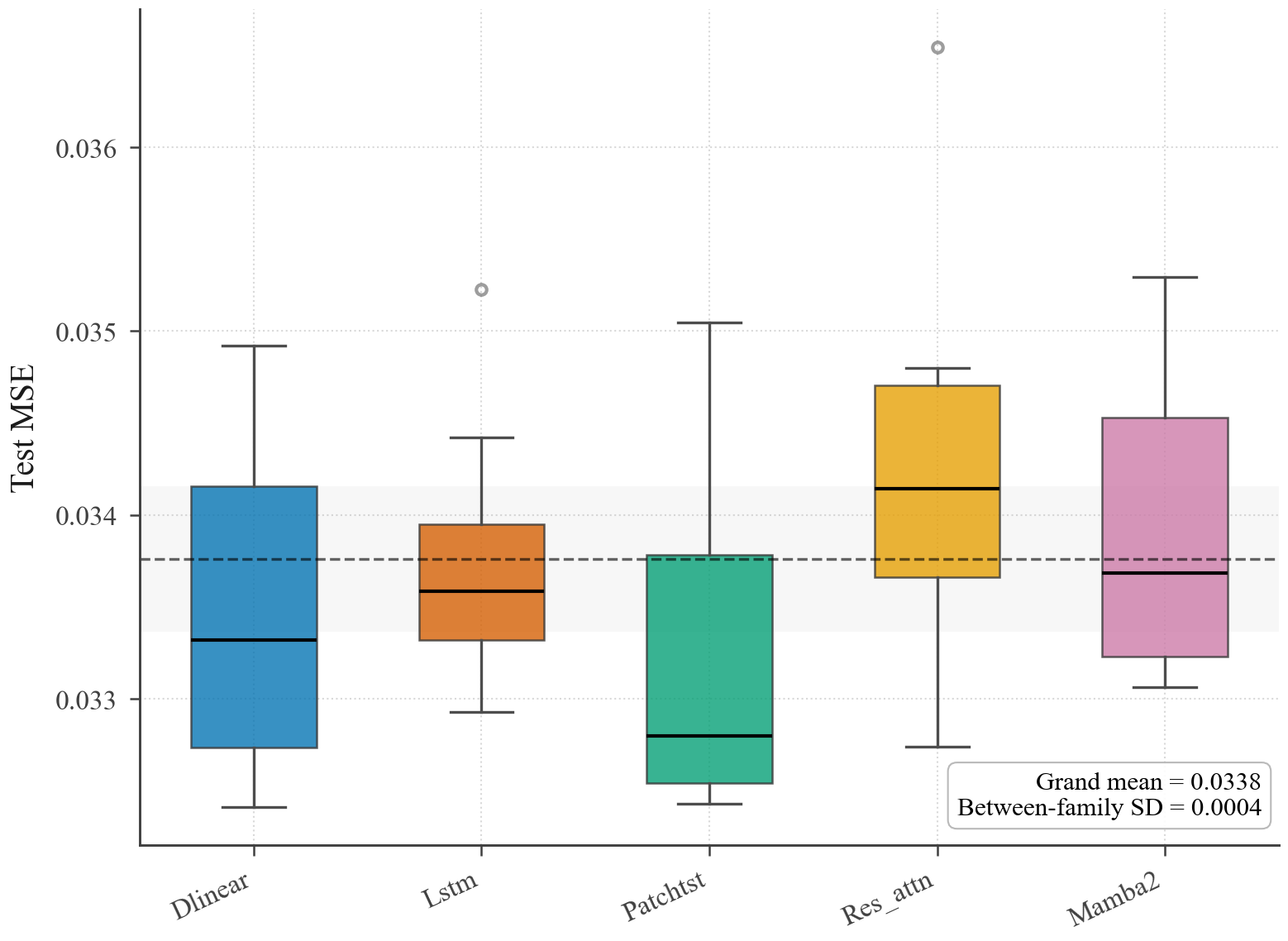}
    \caption{Test MSE distribution across the nine context lengths per family, station 41009. Boxes show median and IQR; whiskers extend to $1.5\times$ IQR. Dashed line: grand mean. Shaded band: $\pm 1$ between-family SD.}
    \label{fig:mse_sweep}
\end{figure}

\begin{figure}[ht]
    \centering
    \begin{subfigure}[t]{0.48\textwidth}
        \centering
        \includegraphics[width=\textwidth]{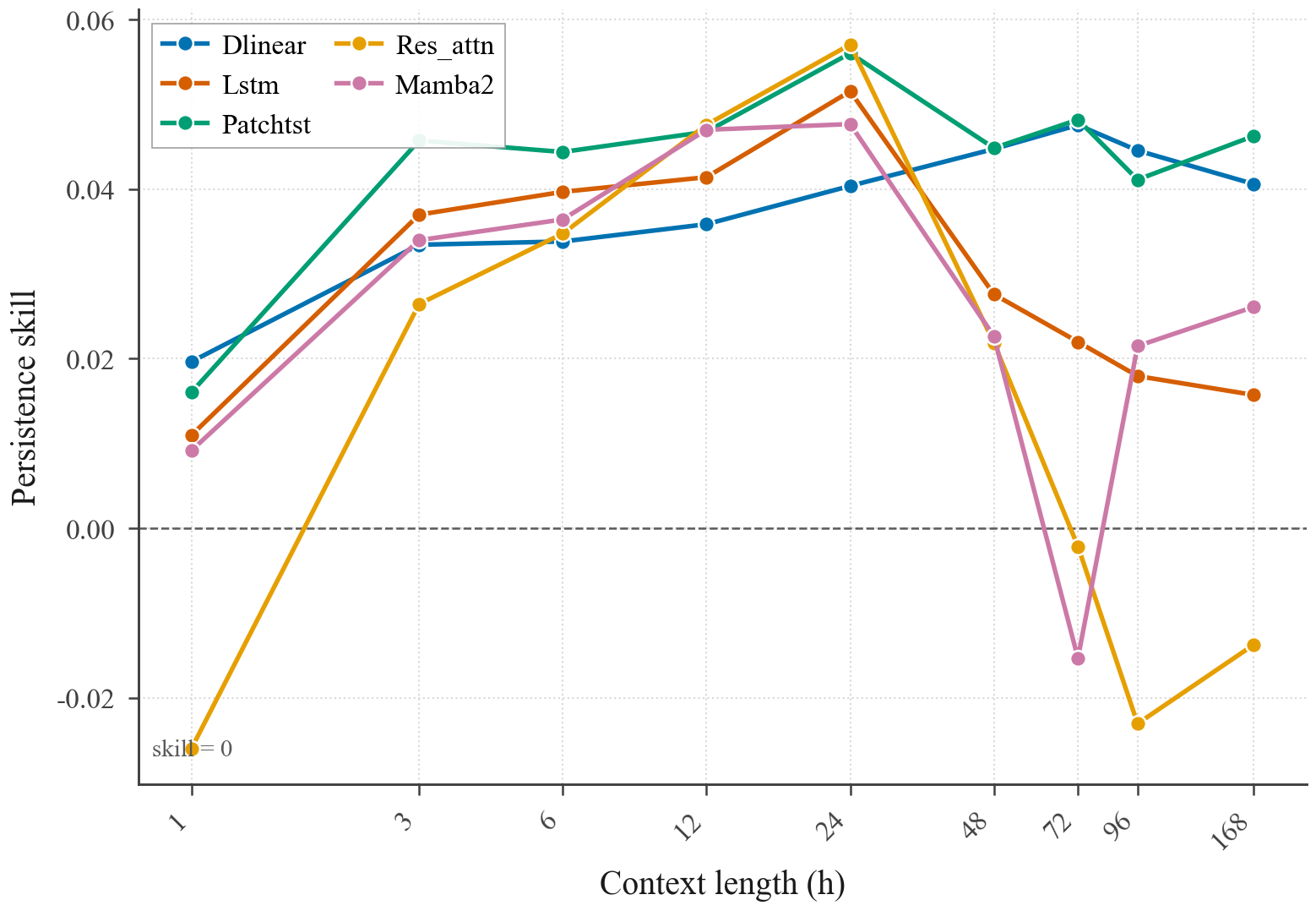}
        \caption{Persistence skill as a function of context length, station 41009.}
        \label{fig:skill_sweep}
    \end{subfigure}
    \hfill
    \begin{subfigure}[t]{0.48\textwidth}
        \centering
        \includegraphics[width=\textwidth]{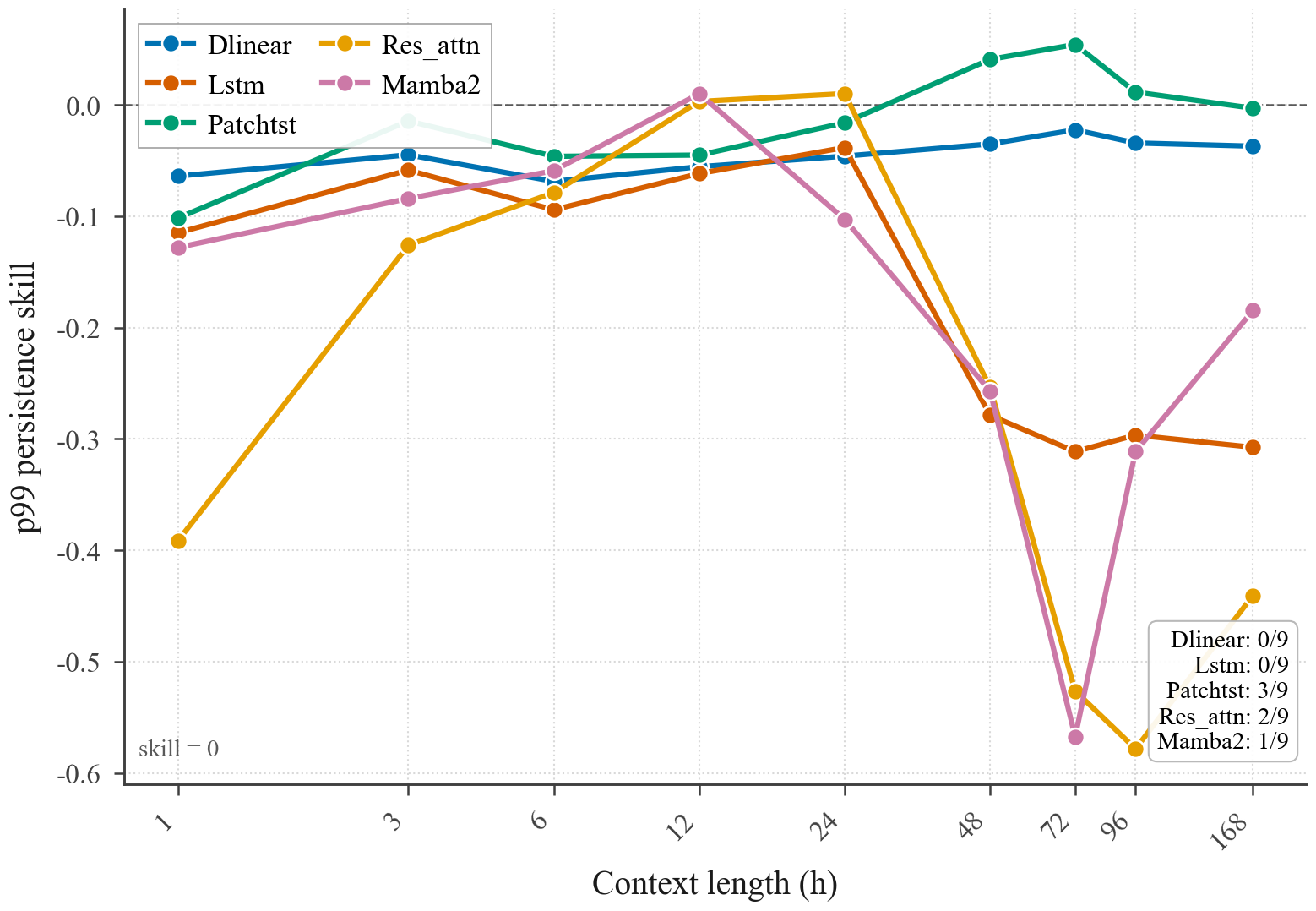}
        \caption{$p_{99}$ tail persistence skill as a function of context length, station 41009. Inset box: number of CLs with positive $p_{99}$ skill per family.}
        \label{fig:p99_skill}
    \end{subfigure}
    \caption{Persistence skill across context lengths, station 41009}
\end{figure}

The 2,700-trial Optuna search on station 41009 produced 45 best configurations (one per family $\times$ CL cell). The complete per-trial metrics for all 90 (family, CL) cells in both the single-buoy and multi-buoy phases are provided as Supplementary Material. The persistence baseline achieved a test MSE of $0.0348$~m\textsuperscript{2} on this dataset. Three diagnostic patterns emerged: (i) all five families converged to nearly identical test MSE (Table~\ref{tab:per_family_results}; Figure~\ref{fig:mse_sweep}), with a between-family standard deviation of only $0.0004$~m\textsuperscript{2}---the MSE range across context lengths ($0.0330$--$0.0354$~m\textsuperscript{2}) exceeded the across-family range ($0.0333$--$0.0343$~m\textsuperscript{2}), indicating that context length accounted for more variance than architecture choice; all 45 trials yielded Pearson $r \geq 0.96$; (ii) persistence skill peaked at $\mathrm{CL} = 12$--24~h (range $+0.036$ to $+0.057$ across the five families) and degraded at longer $\mathrm{CL}$ (Figure~\ref{fig:skill_sweep})---at $\mathrm{CL} = 1$, models added negligible value beyond the last observation, and five of 45 trials exhibited negative skill (four ResAttLstm, one Mamba2); (iii) extreme-wave prediction was uniformly poor: $p_{99}$ MSE exceeded the overall test MSE by a factor of $13$--$16\times$, only 6 of 45 trials achieved positive $p_{99}$ persistence skill (Figure~\ref{fig:p99_skill}), and per-family Q4/$Q3$ test MSE ratios ranged from $2.4$ to $2.6$ (Table~\ref{tab:per_family_results}). The worst individual (family, CL) cells reached Q4/$Q3$ ratios above $3.0$ (ResAttLstm@96h: 3.2; Mamba2@72h: 3.1; LSTM@72h: 3.0).

\begin{table}[ht]
    \centering
    \small
    \begin{tabular}{lcccc}
        \hline
        Family     & Test MSE & Persistence skill             & p99 MSE & Q4/Q3 ratio \\
        \hline
        PatchTST   & 0.0333   & +0.043                        & 0.446   & 2.4         \\
        DLinear    & 0.0335   & +0.038                        & 0.459   & 2.4         \\
        LSTM       & 0.0338   & +0.029                        & 0.516   & 2.5         \\
        Mamba2     & 0.0339   & +0.026 ($-0.015$ -- $+0.048$) & 0.521   & 2.5         \\
        ResAttLstm & 0.0343   & +0.014 ($-0.026$ -- $+0.057$) & 0.558   & 2.6         \\
        \hline
    \end{tabular}
    \caption{Per-family results on station 41009. Mean values across nine context lengths. The overall skill range (min--max across CLs) is shown for the two families with negative-skill trials.}
    \label{tab:per_family_results}
\end{table}

\clearpage
\subsection{At-scale evaluation}
\label{scale_evaluation}

The 45 best configurations from Phase 1 were re-trained and evaluated on the concatenated 47-buoy, 37-year dataset. The multi-buoy persistence baseline test MSE, averaged over the nine context lengths, is $0.1724$~m\textsuperscript{2} (the CL-dependent variation of this baseline is addressed in Section~\ref{sec:limitations}).

\begin{figure}[ht]
    \centering
    \includegraphics[width=0.8\textwidth]{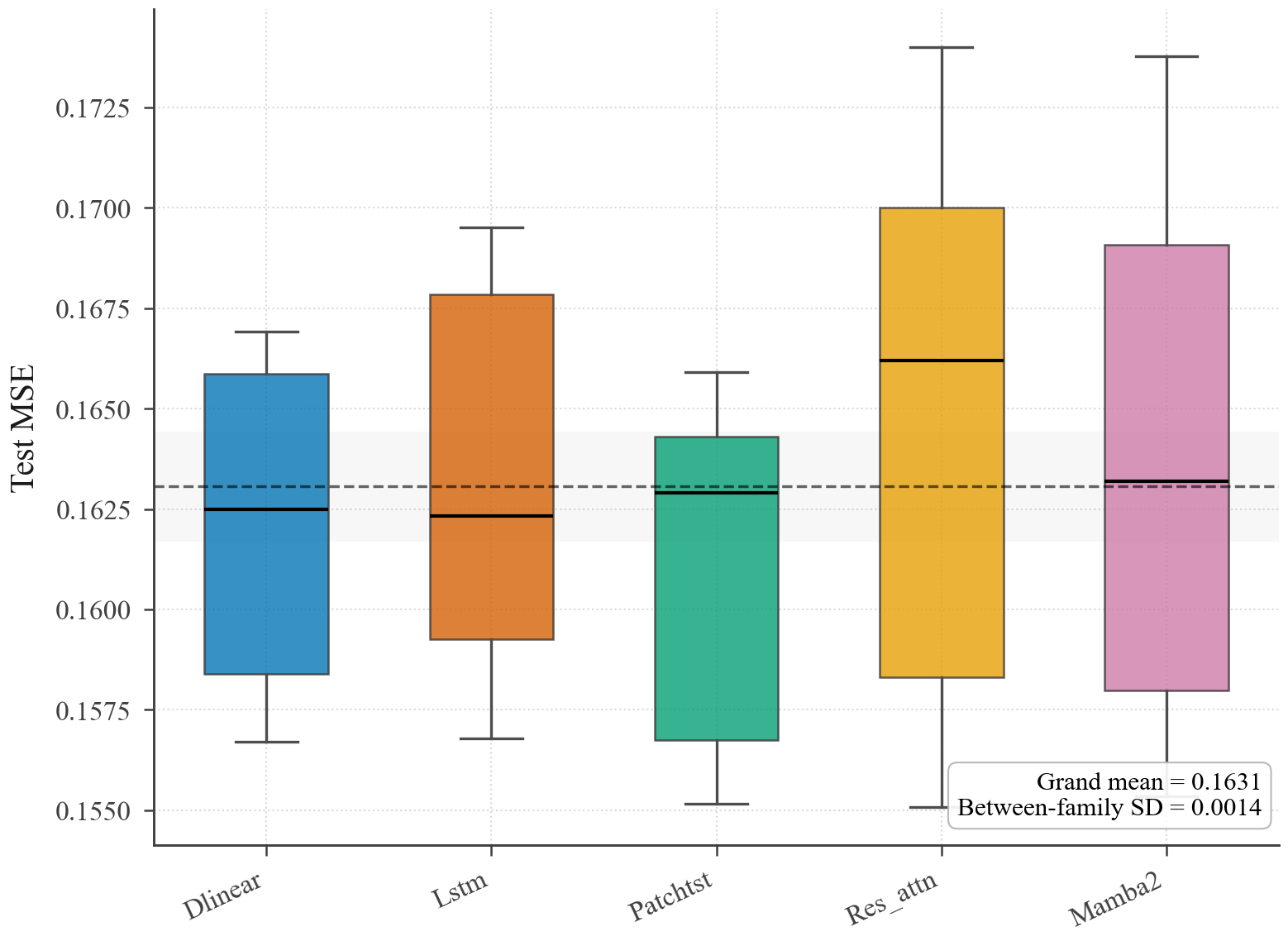}
    \caption{Test MSE distribution across the nine context lengths per family, 47-buoy dataset. Boxes show median and IQR; whiskers extend to $1.5 \times \text{IQR}$.}
    \label{fig:mse_retrain}
\end{figure}

\begin{figure}[ht]
    \centering
    \begin{subfigure}[t]{0.48\textwidth}
        \centering
        \includegraphics[width=\textwidth]{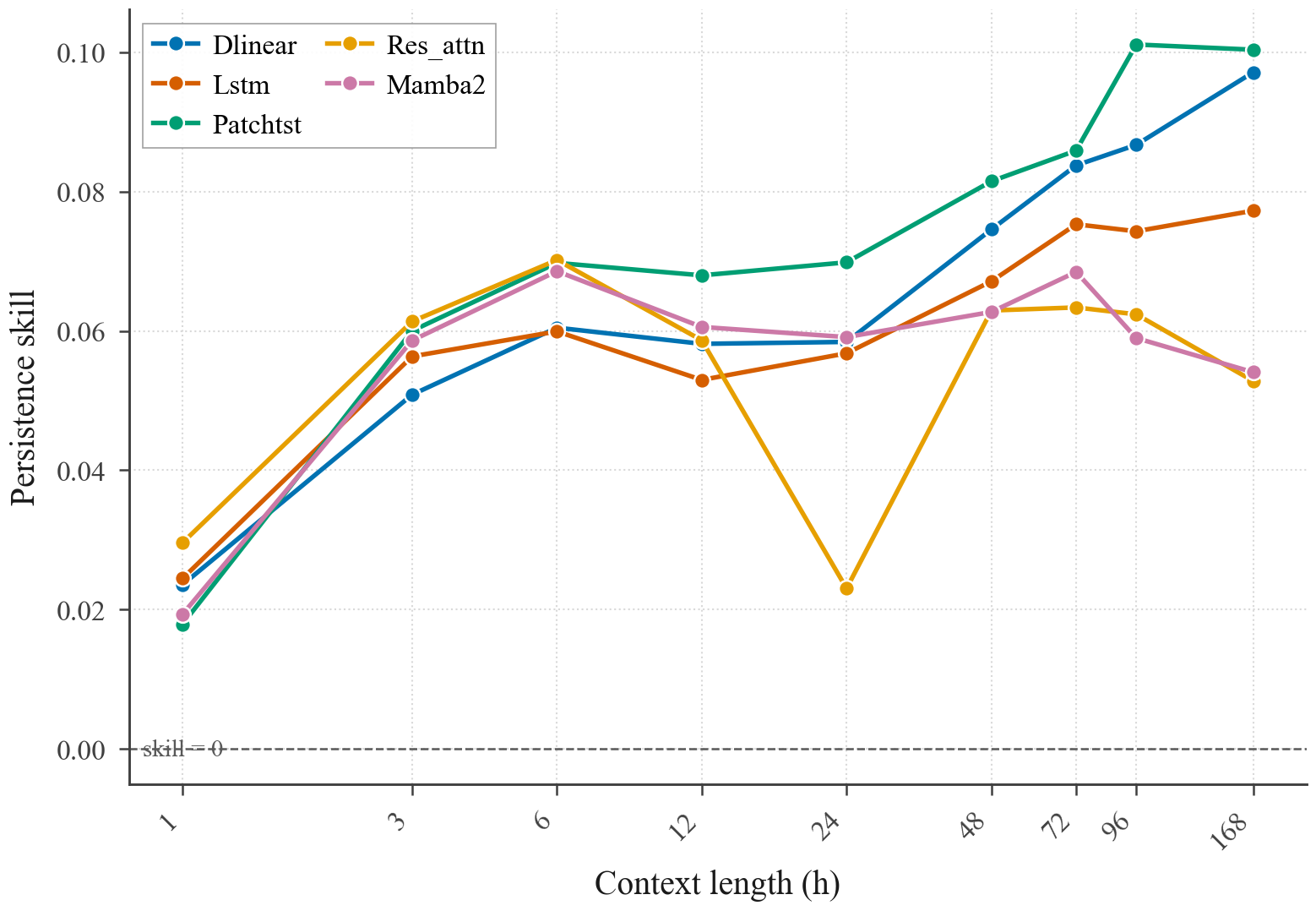}
        \caption{Persistence skill as a function of context length, 47-buoy dataset.}
        \label{fig:skill_retrain}
    \end{subfigure}
    \hfill
    \begin{subfigure}[t]{0.48\textwidth}
        \centering
        \includegraphics[width=\textwidth]{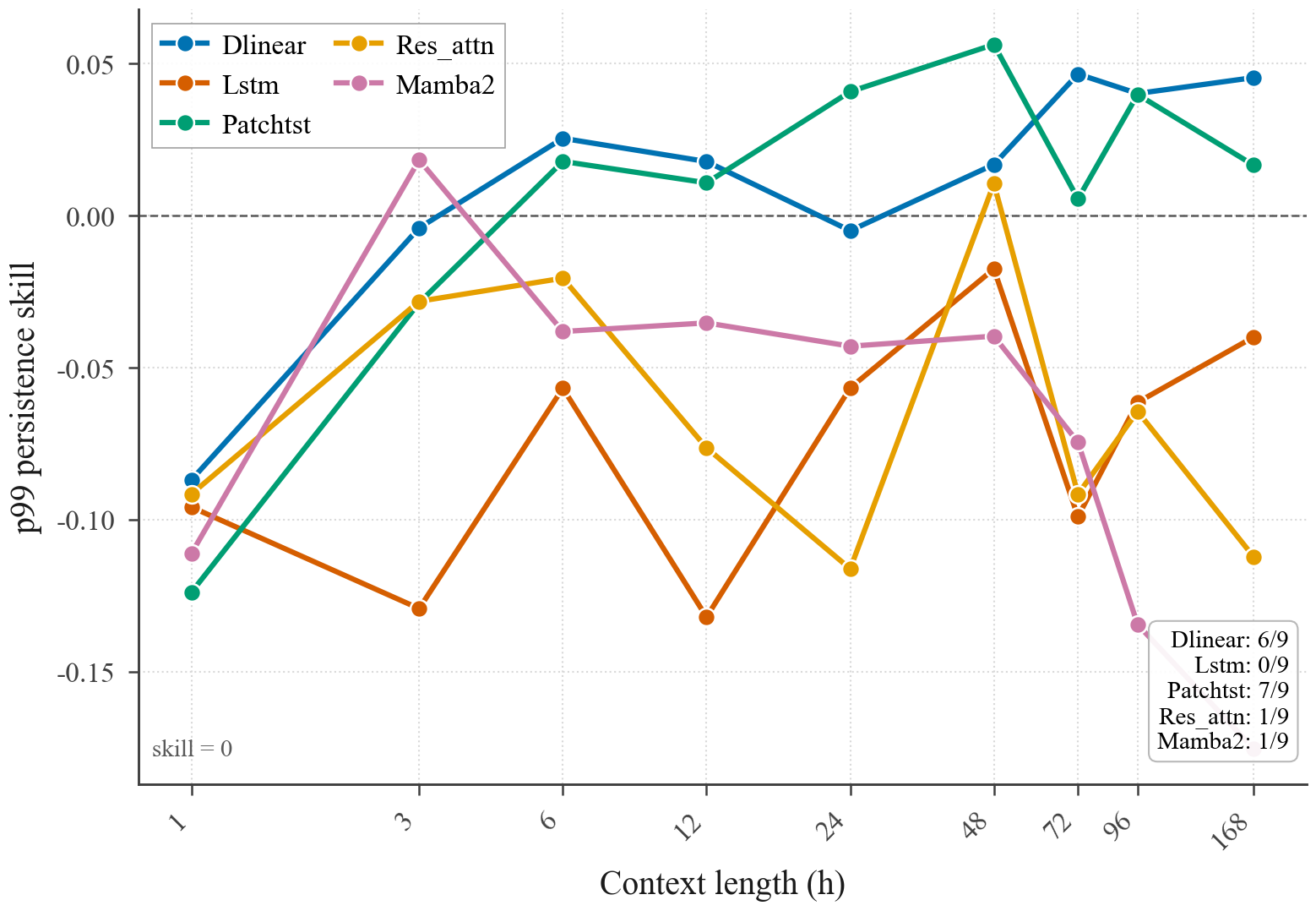}
        \caption{$p_{99}$ tail persistence skill as a function of context length, 47-buoy dataset. Inset box: number of CLs with positive $p_{99}$ skill per family.}
        \label{fig:p99_skill_retrain}
    \end{subfigure}
    \caption{Persistence skill and tail performance on the 47-buoy dataset}
\end{figure}

The five families converge to nearly identical multi-buoy test MSE: the five-family grand mean is $0.1631$~m\textsuperscript{2} with a between-family standard deviation of $0.0014$~m\textsuperscript{2}---only $0.8\%$ of the mean (Figure~\ref{fig:mse_retrain}). All 45 trials achieve positive persistence skill (5-family mean $+0.062$; Figure~\ref{fig:skill_retrain}), and no architecture consistently outperforms the others.

Tail errors are proportionally smaller than on the single-buoy set but the inter-family spread remains similarly small. The $p_{99}$ MSE reaches $1.91$--$2.08$~m\textsuperscript{2} across families---approximately $12\times$ the overall multi-buoy MSE---with an inter-family spread of only 9\% of the mean. The five-family grand mean $p_{99}$ persistence skill is $-0.040$ (Figure~\ref{fig:p99_skill_retrain}): despite improvement over the single-buoy case, deep models still underperform persistence on average for the most extreme 1\% of samples. Only 15 of 45 trials achieve positive $p_{99}$ skill (up from 6/45 on the single-buoy set; per-family breakdown in Figure~\ref{fig:p99_skill_retrain}).

\begin{figure}[ht]
    \centering
    \includegraphics[width=0.8\textwidth]{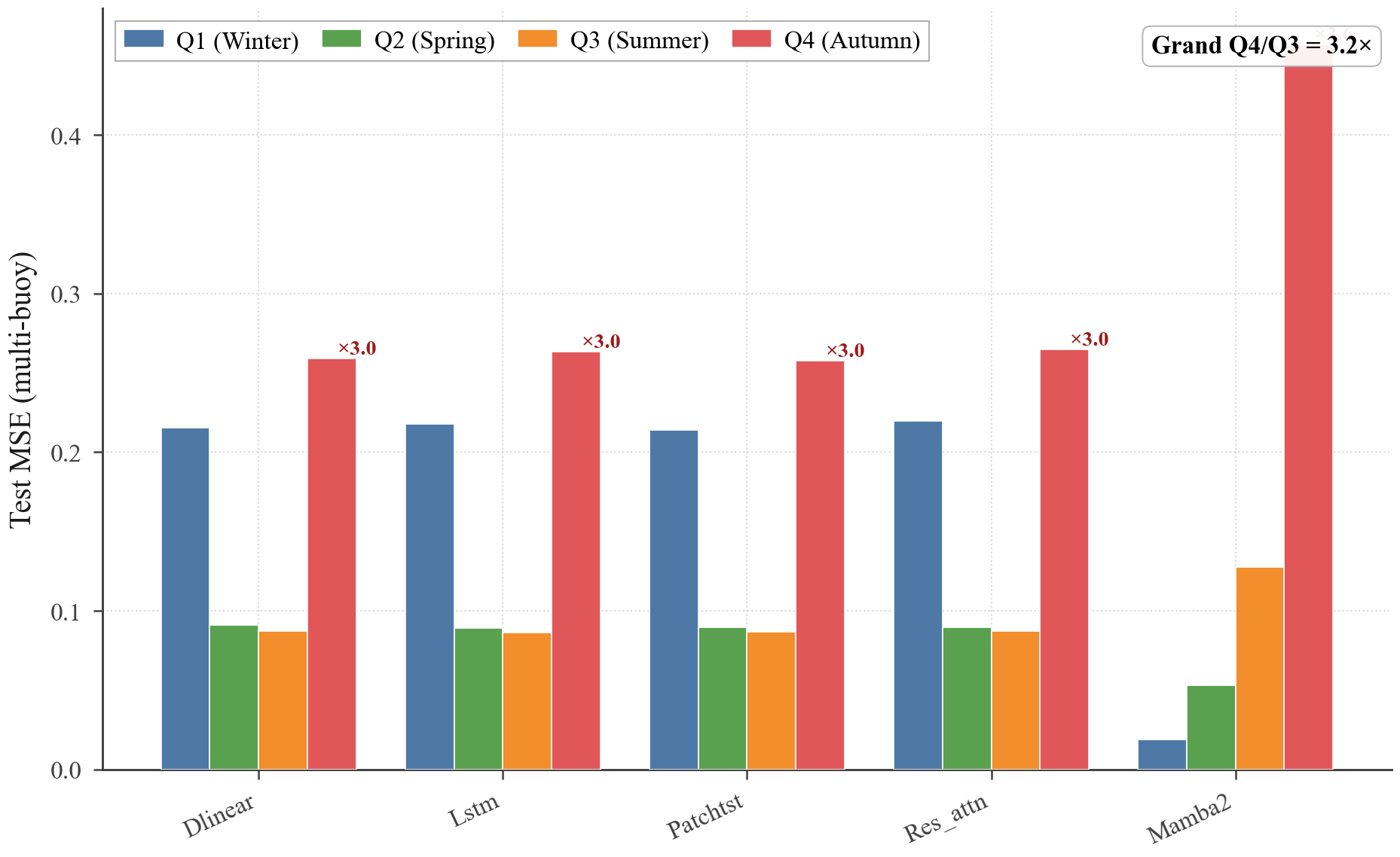}
    \caption{Mean seasonal test MSE by family, 47-buoy dataset. Numbers above Q4 bars: per-family Q4/Q3 MSE ratio.}
    \label{fig:seasonal_mse}
\end{figure}

Seasonally, Q4 (October--December, encompassing the early winter storm season) test MSE exceeds Q3 (July--September, summer) by a factor of $2.8$--$3.2$ across families (Figure~\ref{fig:seasonal_mse}). When normalised by seasonal persistence MSE, $Q4$ skill is consistently below $Q3$ skill: the 5-family grand mean is $Q3 = +0.084$ and $Q4 = +0.055$, with a mean within-trial difference of $\Delta = -0.029$. This pattern holds in 42 of the 45 (family, CL) cells (the three exceptions are all at $\mathrm{CL} = 1$, where seasonal differences are minimal) and across all five families individually (Table~\ref{tab:seasonal_skill}), indicating that the Q4 performance drop is not solely an artefact of higher SWH variance.

\begin{figure}[ht]
    \centering
    \includegraphics[width=0.8\textwidth]{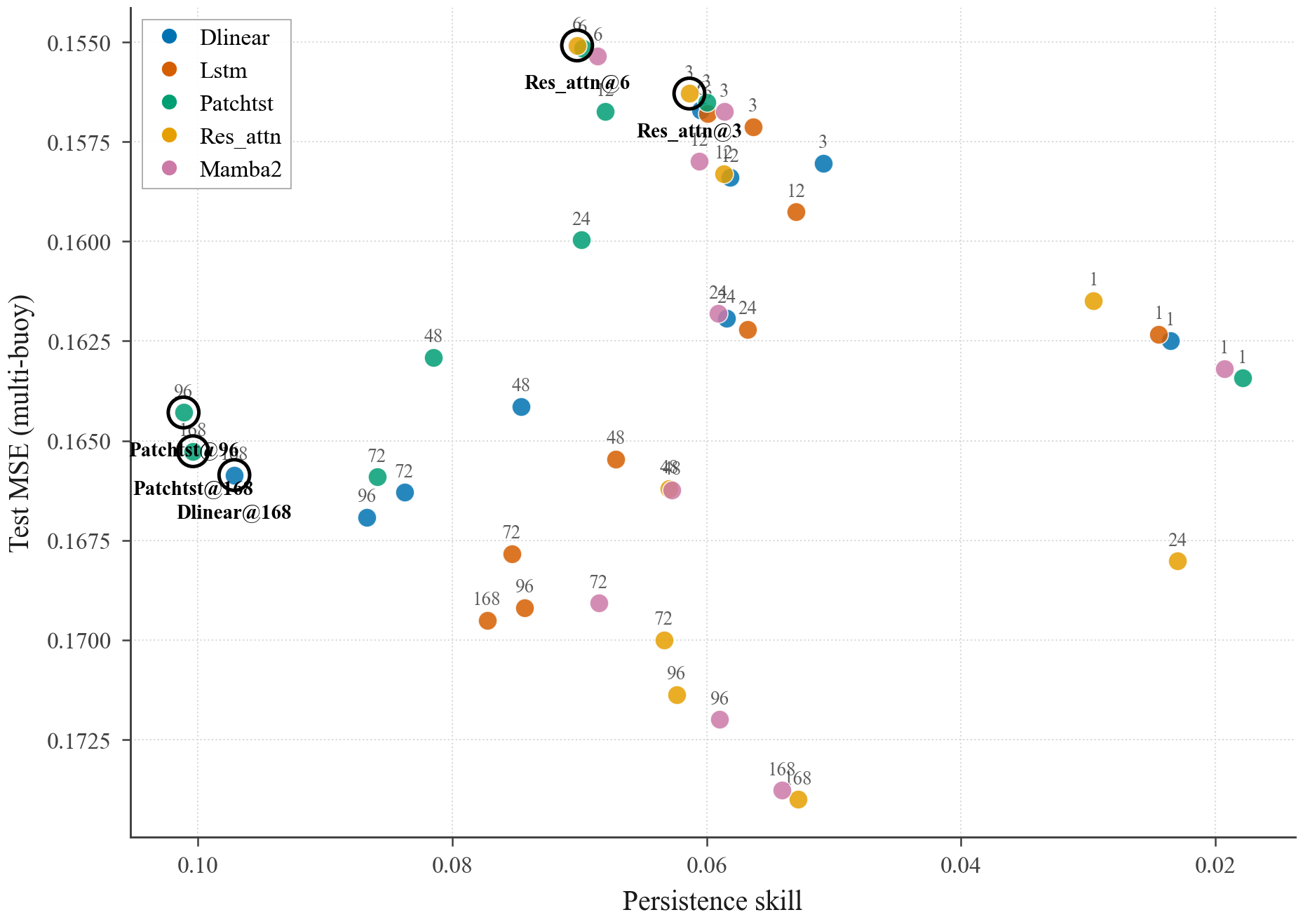}
    \caption{Test MSE versus persistence skill for each (family, context length) cell, 47-buoy dataset. Annotated numbers indicate context length in hours. Circled points mark Pareto-optimal configurations.}
    \label{fig:decision_matrix}
\end{figure}

The interaction between MSE and skill across the 45 (family, CL) cells reveals a compact trade-off space (Figure~\ref{fig:decision_matrix}): ResAttLstm and PatchTST occupy the low-MSE frontier at short CL, while PatchTST and DLinear dominate the high-skill frontier at long CL. No single (family, CL) configuration simultaneously minimises MSE and maximises skill, and the Pareto front is composed primarily of short-CL ResAttLstm/PatchTST and long-CL PatchTST/DLinear configurations.

\begin{table}[ht]
    \centering
    \small
    \resizebox{\textwidth}{!}{%
    \begin{tabular}{lccccc}
        \hline
        Family & Q1 & Q2 & Q3 & Q4 & $\Delta$(Q4$-$Q3) \\
        \hline
        DLinear             & +0.058 & +0.072 & +0.081 & +0.064 & $-0.017$ \\
        LSTM                & +0.047 & +0.091 & +0.090 & +0.050 & $-0.040$ \\
        PatchTST            & +0.064 & +0.084 & +0.087 & +0.070 & $-0.017$ \\
        ResAttLstm          & +0.040 & +0.083 & +0.084 & +0.044 & $-0.040$ \\
        Mamba2              & +0.046 & +0.085 & +0.078 & +0.048 & $-0.030$ \\
        \hline
        \textbf{Grand mean} & \textbf{+0.051} & \textbf{+0.083} & \textbf{+0.084} & \textbf{+0.055} & \textbf{$-0.029$} \\
        \hline
    \end{tabular}%
    }
    \caption{Seasonal persistence skill by family, multi-buoy evaluation. Values are 9-CL means. All five families exhibit Q4 skill $<$ Q3 skill.}
    \label{tab:seasonal_skill}
\end{table}

\clearpage
\subsection{Cross-dataset transfer}
\label{cross_dataset_transfer}

\begin{figure}[ht]
    \centering
    \includegraphics[width=0.8\textwidth]{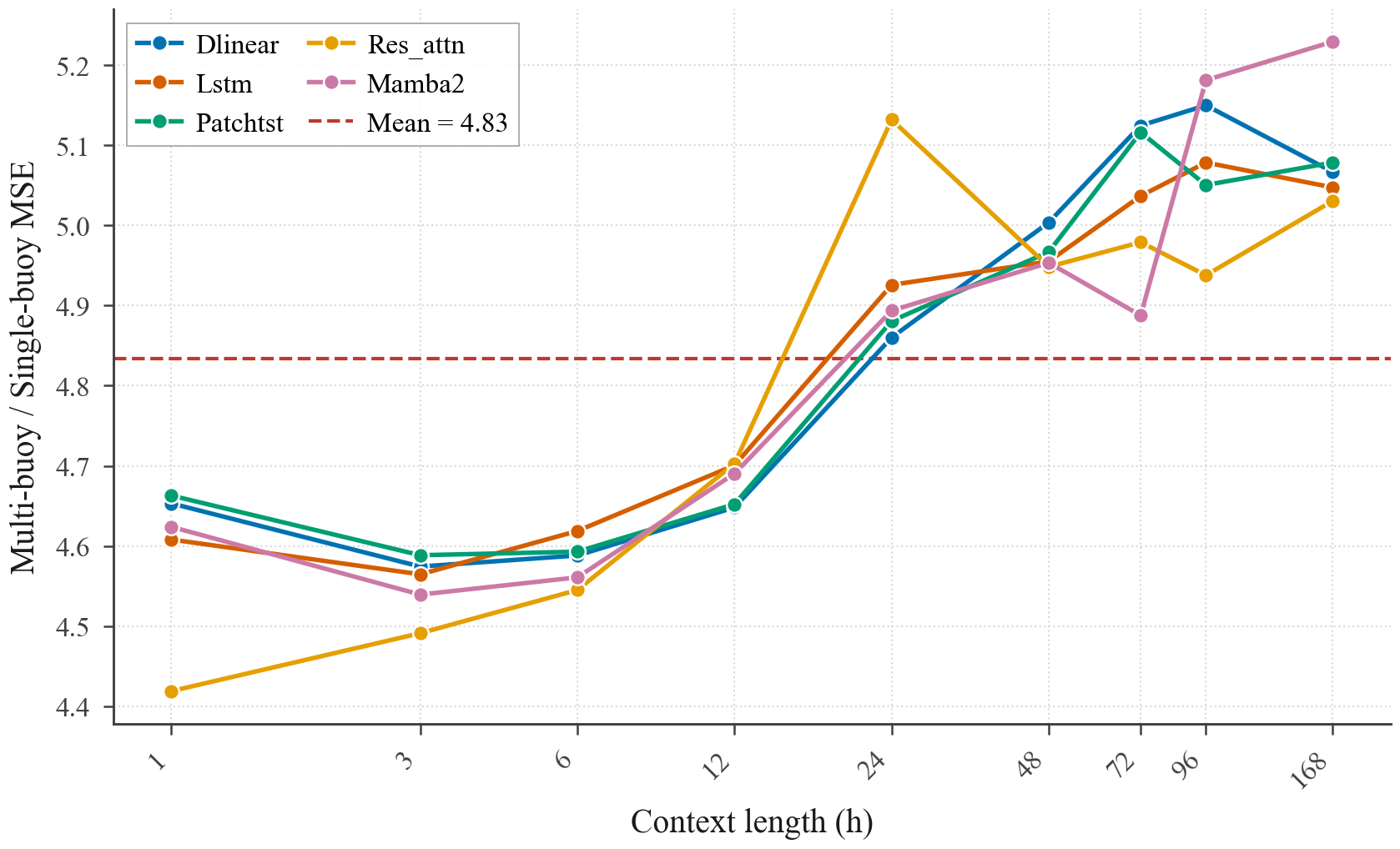}
    \caption{Ratio of multi-buoy to single-buoy test MSE per family and context length. The dashed line marks the grand mean (4.83).}
    \label{fig:mse_scaling}
\end{figure}

The 47-buoy test MSE exceeds the single-buoy test MSE by a factor of 4.26--5.15 (mean 4.83, standard deviation 0.23) across the 45 (family, CL) pairs---an absolute increase of 0.129~m\textsuperscript{2} from the single-buoy grand mean (0.0338~m\textsuperscript{2}) to the multi-buoy grand mean (0.1631~m\textsuperscript{2}), with a bootstrap 95\% CI of $[4.71, 4.87]$ for the ratio. The ratio is larger at long CL (mean 5.09 at CL = 96, 168) than at short CL (mean 4.57 at CL = 1, 3) (Figure~\ref{fig:mse_scaling}). This cross-dataset shift dominates all other sources of variation: the absolute increase of 0.129~m\textsuperscript{2} is approximately $330\times$ the single-buoy between-family standard deviation ($0.0004$~m\textsuperscript{2}) and $95\times$ the multi-buoy between-family standard deviation ($0.0014$~m\textsuperscript{2}).

\begin{figure}[ht]
    \centering
    \includegraphics[width=0.8\textwidth]{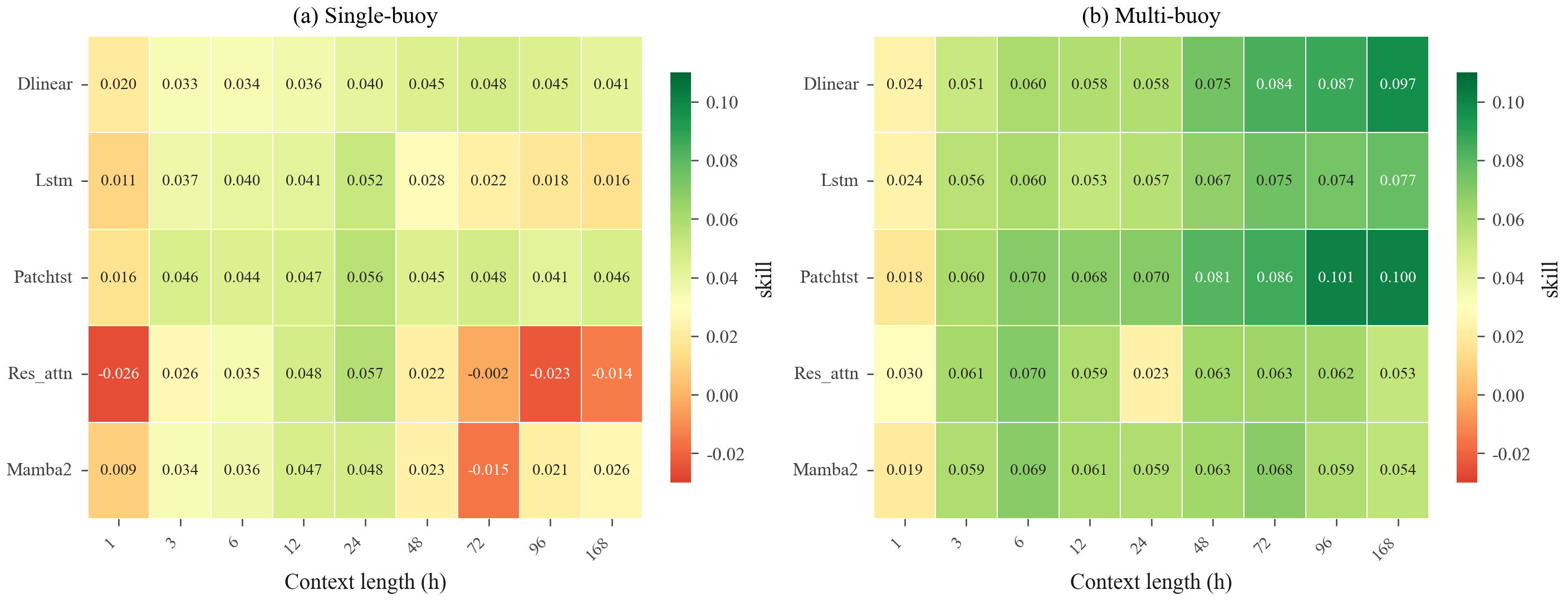}
    \caption{Persistence skill per family and context length. (a) Single-buoy. (b) Multi-buoy.}
    \label{fig:ranking_flip}
\end{figure}

The ranking of (family, CL) pairs by persistence skill reverses between the single-buoy and multi-buoy experiments. LSTM, which ranks among the top families on the single-buoy set, is not the winner at any CL on the multi-buoy set. ResAttLstm@24 drops from first (single-buoy skill $+0.0571$) to fifth ($+0.0230$) at CL = 24 on the multi-buoy set. Mamba2@72 recovers from negative ($-0.0153$) to positive ($+0.0685$), and Mamba2@168 improves from $+0.0261$ to $+0.0541$ (Figure~\ref{fig:ranking_flip}).

\subsection{Robustness to missing observations}
\label{sec:missing_data}

To assess whether the preceding conclusions are sensitive to the data-quality issues common in operational buoy records, 5\% and 15\% of hourly $H_s$ observations were randomly removed from the 47-buoy test set, imputed via linear interpolation, and all 45 best configurations re-evaluated without retraining (full protocol in Section~\ref{sec:practical_scope}). Table~\ref{tab:missing_data} reports the 9-CL mean test MSE and persistence skill per family at each missingness level.

\begin{table}[ht]
    \centering
    \small
    \resizebox{\textwidth}{!}{%
    \begin{tabular}{lcccccc}
        \hline
        Family     & MSE (0\%) & MSE (5\%) & MSE (15\%) & Skill (0\%) & Skill (5\%) & Skill (15\%) \\
        \hline
        PatchTST   & 0.1611   & 0.1601   & 0.1565    & +0.0727     & +0.0730     & +0.0737      \\
        DLinear    & 0.1623   & 0.1613   & 0.1575    & +0.0660     & +0.0663     & +0.0676      \\
        LSTM       & 0.1633   & 0.1624   & 0.1591    & +0.0605     & +0.0600     & +0.0586      \\
        Mamba2     & 0.1640   & 0.1631   & 0.1597    & +0.0567     & +0.0562     & +0.0551      \\
        ResAttLstm & 0.1645   & 0.1636   & 0.1603    & +0.0538     & +0.0532     & +0.0516      \\
        \hline
    \end{tabular}%
    }
    \caption{Per-family 9-CL mean test MSE and persistence skill under 0\%, 5\%, and 15\% randomly missing observations with linear interpolation, multi-buoy evaluation.}
    \label{tab:missing_data}
\end{table}

Three patterns are observed. First, test MSE does not increase under missing data: the five-family mean decreases by $0.0009$~m\textsuperscript{2} ($-0.6\%$) at 5\% missingness and $0.0044$~m\textsuperscript{2} ($-2.7\%$) at 15\% missingness. Second, persistence skill is effectively unchanged: the five-family grand mean skill shifts by $-0.0002$ at 5\% and $-0.0006$ at 15\%, remaining within $\pm 0.003$ of the complete-data baseline for every family at both levels. Third, the inter-family ranking is fully preserved at both missingness levels (PatchTST $>$ DLinear $>$ LSTM $>$ Mamba2 $>$ ResAttLstm, by both MSE and skill).

\clearpage
\section{Discussion}
\label{discussion}
\subsection{Why the five families are of no practical consequence}
\label{sec:negligible}
The five families produce test MSE values that are negligible in practice on both the single-buoy and multi-buoy evaluations. The between-family SD of $0.0014$~m$^2$ on the multi-buoy set corresponds to an RMSE difference of approximately 0.3~cm---comparable to, or smaller than, the measurement precision of typical NDBC wave sensors (which report $H_s$ to 0.1~m resolution), and two orders of magnitude below the 21.8~cm RMSE shift induced by switching buoy corpora. This convergence across linear decomposition, patched attention, CNN-attention-LSTM hybrids, recurrent, and state-space architectures—at nine context lengths and across a 47-buoy, 37-year corpus—indicates that the result is not an artefact of any particular model class or experimental configuration.

The most parsimonious interpretation---offered here as a post-hoc hypothesis, since the experiment was designed to compare architectures rather than to test inertial dynamics---is that the five families are learning the same linear-inertial SWH signal that is already captured by persistence. On short timescales (minutes to a few hours), SWH varies smoothly because the wave field responds to integrated wind forcing over many wave periods; under this interpretation, a one-step-ahead prediction of $H_s$ from past $H_s$ alone is, on these horizons, an inertial forecast: the best prediction is the last observed value, and any model that deviates from this prediction can at best learn a correction on the order of the local measurement error. If this interpretation is correct, persistence is already near-optimal for short-horizon $H_s$, and the deep models are fitting residual noise around the inertial signal. The non-linear correction on top of this inertial signal, if present, accounts for a small fraction of the total variance, and the choice of architecture does not materially affect its magnitude---consistent with the observed between-family SD of $0.0004$~m$^2$ (Table~\ref{tab:per_family_results}). This interpretation makes a testable prediction: if models are primarily fitting residual noise, the skill gain over persistence should be uncorrelated with context length, since noise lacks temporal structure. The weak but positive correlation between skill and CL observed in the multi-buoy experiment (Figure~\ref{fig:skill_retrain}) is therefore not fully explained by the noise-fitting account alone, suggesting that a small structured signal is also being extracted.

A complementary physical hypothesis for why univariate $H_s$ may impose a ceiling follows from the two-component structure of the wave field \citep{holthuijsen2007waves}. Wind-sea---waves generated by local wind---responds to changes in wind speed and direction on timescales of hours; if this evolution is not encoded in past $H_s$ values, it cannot be forecast from $H_s$ alone beyond the inertial timescale of the existing wave field. Swell---waves that have propagated away from their generation region---evolves over days and could, in principle, carry a longer memory extractable by a univariate model. However, without access to the wind field that generated the swell, nor to the propagation distance or directional spread, any extractable signal would be weak and would saturate quickly. Under this two-component hypothesis, the wind-sea component becomes unpredictable from $H_s$ alone below approximately 24~h, while the swell component would provide a small, architecture-independent, and slowly saturating residual signal. The linear-inertial and two-component hypotheses are not mutually exclusive---both predict architecture independence and skill saturation---and the present experiment cannot distinguish between them. Disentangling these mechanisms would require explicit decomposition of $H_s$ into wind-sea and swell components, which is left to future work.

Beyond temporal wind-sea/swell partitioning, real wave fields in harbours and near marine structures involve inherently spatial mechanisms---harbour resonance, Bragg reflection over periodic bathymetry, and gap resonance between closely spaced floating bodies---that cannot be encoded in a single-point $H_s$ time series. These processes have been extensively characterised through phase-resolving numerical models \citep{gao2020numerical,gao2021investigation,gao2023mechanism,gao2024influences,gong2024hydrodynamics,mi2025gap,gao2026influences,gao2026hydrodynamics}, yet none of the spatial information they capture---wave direction, basin geometry, inter-body spacing---is available to a univariate forecaster. This spatial information deficit represents a second, independent ceiling on forecast skill that no architecture alone can overcome, consistent with the observation that all five families converge to the same MSE regardless of their capacity to model complex temporal dependencies.

\subsection{Context-length scaling}

We refer to the central single-buoy pattern as the ``6-hour barrier'': the marginal gain in persistence skill per additional hour of context drops sharply after approximately 6~h and turns negative beyond 24~h on station 41009 (Figure~\ref{fig:skill_sweep}), and the 2,700-trial search found no configuration that reverses this trend. The skill peak at 12--24~h reflects the optimal trade-off between the small residual benefit of moderate context and the noise accumulated from longer, less informative history. This barrier is a finding specific to the single-buoy experiment; it does not necessarily imply a hard limit on univariate $H_s$ forecasting in general.

On the multi-buoy evaluation, the relationship is more complex and does not exhibit the monotonic degradation seen on the single buoy. Crucially, the ``6-hour wall'' does not hold in this setting. Mean skill rises from CL = 1 to CL = 6 (+0.023 to +0.066), dips at $\text{CL} = 12$-$24$ (+0.060 and +0.053), then rises from $\text{CL} = 48$ to $\text{CL} = 96$ (+0.070 to +0.077) and plateaus through $\text{CL} = 168$ (+0.076). The best single-trial skill is +0.1011 at $\text{CL} = 96$~h (PatchTST, Figure~\ref{fig:skill_retrain}). This divergence from the single-buoy pattern---where skill peaks at 12--24~h and degrades beyond---is consistent with the $4.83\times$ cross-dataset MSE shift (Section~\ref{cross_dataset_transfer}) and the ranking reversal between phases (Figure~\ref{fig:ranking_flip}): the heterogeneous 47-buoy corpus presents a materially different forecasting problem from station 41009 alone, and context-length trends observed on a single buoy do not transfer to the multi-buoy setting. Two physical mechanisms may explain why a larger buoy corpus breaks the 6-hour barrier observed on a single station. First, swell propagation across the buoy network provides long-range temporal memory: a swell event recorded at one buoy may, after a lag of hours to days, appear at another buoy downstream, and models trained on the pooled corpus can exploit these cross-buoy propagation delays as implicit long-range features. Second, buoy mixing alters the effective sample distribution: pooling buoys from diverse wave climates broadens the training distribution, potentially shifting the optimal context window away from the 12--24~h peak observed on station 41009 alone. These hypotheses remain to be tested through controlled ablation of buoy subsets. The inter-family skill spread increases from 0.010 at $\text{CL} = 6$ to 0.048 at $\text{CL} = 168$, but this widening is driven primarily by ResAttLstm, which drops from best-at-6~h at +0.070 to worst-at-168~h at +0.053; the four remaining families fall within a 0.046 range at $\text{CL} = 168$ (DLinear +0.097, PatchTST +0.100, LSTM +0.077, Mamba2 +0.054). In absolute terms, this 0.046 skill spread corresponds to an RMSE difference of approximately 0.4~cm---two orders of magnitude smaller than the 21.8~cm RMSE shift between the single-buoy and multi-buoy evaluations---indicating that, even at long context, architecture choice influences the forecast by an amount that is negligible relative to the choice of buoy corpus. The simpler DLinear matches the best-performing PatchTST, and the observation that a linear decomposition model benefits from long context as much as a state-space or patched-attention architecture is consistent with the hypothesis that the learnable signal at long horizons remains predominantly linear-inertial.

On the single-buoy set, per-family mean skill at 1--6~h context ranges from +0.01 to +0.04; on the multi-buoy evaluation, the best-at-6~h reaches +0.0702 (ResAttLstm). Both are consistent with the 5--20\% gains reported in prior SWH studies. The barrier is not a hyperparameter problem---2,700 Optuna trials did not find a configuration that breaks it---nor a model-class problem, as all five inductive biases hit the same ceiling. The best-observed skill of $+0.10$ ($R^2 \approx 0.91$, computed on the multi-buoy test set after inverse transformation to metres) remains substantially below what operational physics-based models achieve with atmospheric forcing: \citet{campos2024development} report $R^2$ values of $0.95$--$0.96$ for the NOAA global wave ensemble at Day~1 with full atmospheric forcing, corresponding to an order-of-magnitude lower MSE. This gap underscores the information deficit of the univariate formulation discussed in Section~\ref{sec:negligible}.

\subsection{Limitations and generalisability}
\label{sec:limitations}

The ranking reversal between the single-buoy and multi-buoy experiments (Section~\ref{cross_dataset_transfer}) indicates that the TPE optimiser exploited idiosyncrasies of station 41009 that do not transfer to the broader buoy corpus. The single-buoy findings thus replicate in direction---all architectures beat persistence on the multi-buoy set, and skill remains within the 5--20\% range reported in prior work---but not in ranking, as the architecture that performs best on 41009 is not the best across 47~buoys. Single-buoy hyperparameter search, while methodologically clean, selects configurations tuned to a specific wave climate; the multi-buoy evaluation reveals a more representative performance ordering. Future controlled comparisons should adopt multi-buoy validation as the primary model-selection criterion.

Cross-context-length skill comparisons should be interpreted with the following caveat. All retrain trials share the same concatenated 47-buoy dataset and the same chronological train-validation-test split. However, because the sliding-window construction discards segments shorter than CL + 6 h, the set of valid test windows is a subset of the test time range that depends on CL: $\text{CL} = 168$ requires 174 consecutive hours with no gaps, whereas $\text{CL} = 6$ requires only twelve. Consequently, the test windows at longer CL are a subset of those at shorter CL, and the subset may differ in composition. The observed variation of persistence MSE across CL (0.1670 at 6-24 h vs. 0.1816 at 72-168 h) reflects this subsetting effect rather than any change in the intrinsic difficulty of persistence; with a missing-data rate below 5\%, the overlap is high and the effect is expected to be modest. Nevertheless, future work should quantify window overlap across CL and, where feasible, use imputation to ensure identical test windows for all context lengths.

A second limitation concerns computational budget: a full multi-buoy cross-validated hyperparameter search, while statistically preferable, was not feasible given the 2,700-trial scale already undertaken. The single-buoy search followed by multi-buoy re-evaluation represents a pragmatic compromise, though it likely underestimates the best-attainable skill on the full corpus. Relatedly, best-trial selection inflates apparent gains relative to the median trial within each (family, CL) cell; future controlled comparisons should report the median skill over many trials rather than the single best trial to provide a more representative estimate of expected performance.

Beyond budget constraints, all models exhibit uniformly poor performance on extreme SWH events (Sections~\ref{sec:hp_search}--\ref{scale_evaluation}). The heavy-tailed nature of the SWH distribution and the sensitivity of extreme waves to wave--current interactions suggest that univariate models lack the physical forcing information needed to anticipate rare events. This reflects both a statistical challenge---extreme events constitute a small fraction of the training distribution---and a methodological one: the log-MSE loss function treats all samples equally and provides no incentive for the model to allocate capacity to the tail. Weighted loss functions, extreme-value theory, or dedicated tail-modelling heads may be necessary for operational forecasting where the largest waves are of primary interest. More broadly, an alternative interpretation of the architectural convergence reported here is that the log-MSE loss, by compressing the heavy tail of the SWH distribution, drives all architectures toward a similar mean-regression solution, potentially masking genuine architectural differences that might emerge under a different training objective. The architectural negligible in practice differences observed in this study are therefore conditional on the log-MSE training criterion.

An additional caveat concerns the aggregation of metrics over lead times. All metrics reported here are averages over the six lead times (1--6~h). It remains possible that architectural differences manifest at specific lead times---for instance, the relative advantage of recurrent architectures at short lead times versus attention-based models at longer horizons---but are obscured by averaging; lead-time-disaggregated analysis is left to future work.

A further concern is the seasonal degradation observed across all families (Section~\ref{scale_evaluation}): models lose discriminative power in autumn and winter, precisely when forecasts are most needed for storm warning. While part of this effect is attributable to higher SWH variance in winter months, the persistence-normalised skill gap indicates a structural weakness. This seasonal fragility is a direct consequence of the univariate formulation: winter wave fields in the NDBC domain are dominated by extratropical cyclones whose evolution cannot be inferred from local $H_s$ history alone.

The present study also deliberately omits atmospheric covariates (wind, pressure, fetch). This omission defines the lower bound of what pure time-series models can achieve and establishes a reference against which future physics-informed or multivariate SWH models can be calibrated. It remains to be demonstrated whether the integration of even a single wind-speed channel lifts the 6-hour barrier identified here. Recent work on physics-informed neural networks for wave-field reconstruction \citep{wang2022reconstruction} and on integrating data-driven with physics-based wave modelling \citep{wang2022integration} suggests that embedding physical constraints into learned representations may narrow the gap between pure data-driven forecasts and operational physics-based systems.

\subsection{Geographic and dynamical scope}
\label{sec:geo_scope}
All experiments in this study are based on NDBC buoy 41009 and the US coastal buoy network, with stations predominantly located in mid-latitude, storm-dominated or swell-mixed environments where tidal forcing is negligible and shallow-water effects are minimal. The findings reported here should not be extrapolated to the following regimes without dedicated validation.

\textit{Low-latitude non-storm regions.} In tropical and equatorial waters, wave fields are typically dominated by narrow-banded, long-period swell with low seasonal variability. Under these conditions, even simple persistence or climatological forecasts can achieve skill levels close to those of complex architectures, and the incremental benefit of deep learning over linear baselines is expected to be even smaller than reported here.

\textit{Semi-enclosed bays and strong-tide estuaries.} These environments feature shallow-water deformation, tidal modulation of wave parameters, bottom friction, depth-induced wave breaking, refraction, and wave--current interaction---all of which produce strong non-linear signatures that cannot be encoded in a single-point $H_s$ time series. The univariate formulation evaluated in this study is ill-suited to such settings; dedicated approaches incorporating bathymetry, tidal harmonics, and spatially resolved forcing are required.

\textit{Polar marginal ice zones.} Where sea ice attenuates wave energy, the remnant wave field is dominated by low-energy components whose predictability from $H_s$ alone is limited. Complex non-linear models are unlikely to yield marked gains over simpler alternatives in these regimes.

Within the mid-latitude storm/swell-dominated environments for which the NDBC network is representative, the present findings provide a practical reference for rapid wave prediction and hazard mitigation. Validation in the regimes listed above, using more diverse datasets and multi-physics inputs, is left to future work.

\subsection{Practical and methodological scope}
\label{sec:practical_scope}
In operational practice, NDBC buoys routinely experience data loss from sensor failure, power outages, biofouling, and communication disruptions during severe weather. Irregular sampling intervals, instrument drift, and extended gaps of hours to days are common. These real-world data-quality issues are not modelled in the present study. The conclusions drawn here are therefore conditional on the availability of complete, regularly sampled input sequences. Future work should extend the missing-data stress test of Section~\ref{sec:missing_data} to irregular sampling and burst-dropout patterns, investigate imputation strategies beyond linear interpolation that preserve forecast skill under data degradation, and develop training objectives resilient to irregular sampling.

The missing-data experiment of Section~\ref{sec:missing_data} carries an implication that extends beyond practical preprocessing guidance: it provides a further line of evidence that the skill ceiling is set by the information content of the univariate input, not by the inductive bias of the forecaster. Removing up to 15\% of the test observations and replacing them with linear interpolates amounts to low-pass-filtering the input signal---attenuating the high-frequency fluctuations that constitute the irreducible noise floor for any model that sees only past $H_s$. That this operation leaves the inter-family ranking unchanged and shifts the grand mean skill by less than $0.001$ is exactly what one would expect if all five architectures are fitting the same linear-inertial signal already captured by persistence: smoothing the residual noise benefits all models equally, and none possesses a mechanism to recover the lost high-frequency components from the univariate history alone. This null result under input degradation is therefore not a negative finding but a positive corroboration of the central thesis: architecture engineering, even under realistic data intermittency, does not materially alter forecast quality for univariate $H_s$.

The five architectures evaluated---DLinear, LSTM, PatchTST, ResAttLstm, and Mamba2---span linear decomposition, recurrent, patched attention, CNN-attention-LSTM hybrid, and state-space paradigms. Several classic architectures, including multilayer perceptrons (MLP) and multi-scale convolutional models such as TimesNet, were not included in the controlled comparison. The selection was guided by the need to cover distinct inductive biases under a common hyperparameter optimisation budget, rather than to exhaustively benchmark all published time series architectures. The convergence observed here is specific to the univariate $H_s$ task on NDBC buoy data and should not be interpreted as a general statement about the relative merits of these architectures on other benchmarks or in other application domains.

The central finding---that five architecture families yield performance differences of no practical significance---is therefore conditional on (i)~the univariate input space ($H_s$ only), (ii)~the per-buoy, single-station forecasting setup, (iii)~the MSE evaluation criterion, and (iv)~the specific preprocessing pipeline (log-transform, robust scaling). Under a multivariate formulation incorporating wind, pressure, or spectral partitions, or under a different training objective (e.g., tail-weighted or quantile loss), the relative ranking of architectures could differ. Whether richer input spaces would reveal genuine architectural advantages remains an open question.

\section{Conclusion}
\label{sec:conclusion}
A 2,700-trial hyperparameter search on NDBC station 41009 across five inductive biases---linear decomposition, recurrence, patched attention, CNN-attention-LSTM hybrids, and state-space models---and nine context lengths, followed by at-scale re-evaluation of the 45 best configurations on 47~buoys spanning 37~years, demonstrates that architecture choice is not the principal determinant of univariate $H_s$ forecast quality. The five families produce test MSE values on the multi-buoy evaluation that are of no practical consequence, with a between-family SD of $0.0014$~m\textsuperscript{2} (0.8\% of the grand mean) that is dwarfed by the $4.83\times$ cross-dataset MSE shift---an absolute increase of $0.129$~m\textsuperscript{2}, approximately $95\times$ the between-family SD. All 45 multi-buoy trials beat persistence (mean skill $+0.062$), but no architecture consistently outperforms the others. The best-observed skill reaches $+0.10$ (PatchTST at $\text{CL}=96$~h), while the five-family mean at $\text{CL}=168$~h is $+0.076$. A supplementary stress test with up to 15\% randomly missing observations and linear interpolation leaves the inter-family ranking unchanged and shifts the grand mean skill by less than $0.001$, corroborating that the information bottleneck lies in the univariate input, not in model capacity. Three negative findings complete the picture: deep models underperform persistence for the most extreme 1\% of waves (grand mean $p_{99}$ skill $-0.040$; only 15 of 45 trials positive); Q4 (October--December, encompassing the early winter storm season) test MSE exceeds Q3 (July--September, summer) by a factor of $2.8$--$3.2\times$ across all families; and the ranking of (family, CL) pairs by skill reverses between the single-buoy and multi-buoy experiments, indicating that single-buoy hyperparameter selection does not transfer to heterogeneous wave climates.

These results indicate that architecture engineering for pure time-series $H_s$ under the univariate input setting evaluated here has reached diminishing returns. A plausible interpretation, consistent with the observed data, is that persistence captures the dominant linear-inertial signal; deep architectures add a small but real correction that is positive across all multi-buoy trials and improves modestly with longer context, yet remains an order of magnitude smaller than the variance introduced by switching buoy corpora. The ``6-hour barrier''---the sharp drop in marginal skill gain per additional hour of context after approximately 6~h---is observed in the single-buoy experiment and should be understood as a finding from that setting. On the multi-buoy evaluation, the pattern is more complex (a dip at 12--24~h followed by modest recovery at 48--168~h); the barrier does not generalise to the 47-buoy corpus in its original form, and the term ``wall'' is not applied to the multi-buoy conclusions. Only DLinear and PatchTST realise a skill gain exceeding $+0.02$ from 6~h to 168~h ($+0.037$ and $+0.031$, respectively); LSTM shows a marginal gain ($+0.017$), while ResAttLstm and Mamba2 degrade slightly, and even architectures purpose-built for long-sequence modelling (PatchTST, Mamba2) do not substantially exceed the best-observed ceiling. This convergence across five inductive biases and nine context lengths indicates that the ceiling is a property of the task---univariate $H_s$ regression on mid-latitude NDBC buoy data---rather than of any particular model class.

Future AI-based SWH research should therefore redirect effort from univariate architecture search toward four directions grounded in the physical and empirical limitations identified here: (i)~integrating atmospheric covariates (wind, pressure, fetch) to supply the forcing information absent from $H_s$ alone, addressing both the skill ceiling and the seasonal fragility observed in autumn and winter; (ii)~zero-shot cross-buoy transfer to assess whether models can generalise without per-buoy retraining, given the ranking reversals observed between phases; (iii)~decomposing $H_s$ into swell and wind-sea components to expose the physical drivers currently invisible to univariate models; and (iv)~developing tail-aware training objectives---weighted loss functions or extreme-value-theoretic heads---to address the systematic underperformance on the largest 1\% of waves, where deep models currently fall below persistence.

\section{Acknowledgments}
\label{sec:acknowledgments}
The author(s) declare financial support was received for the research, authorship, and/or publication of this article. This research was supported by the Shandong Provincial Lab and Talent Program ("Double Hundred Plan for Oversees Experts" Talent Category, Grant No. WSR2024073 and No. WSR2023026), and by the Innovation Project for Graduate Students of Ludong University (Grant No. IPGS2026-103). The authors would like to thank the anonymous reviewers for their valuable feedback and suggestions, which greatly improved the quality of this paper. We also acknowledge the support of the research community and institutions that provided access to the datasets used in this study.

\noindent\textbf{Declaration of generative AI and AI-assisted technologies in the manuscript preparation process.} During the preparation of this work, the authors used GPT Image~2.0 to generate the conceptual schematic figures (Figures~1 and~3). These figures contain no scientific data or experimental results. No AI tools were used for data analysis, statistical computation, or the drafting of scientific conclusions. The authors reviewed and edited all AI-generated visual content and take full responsibility for the final manuscript.

\section{Author Contributions}
\label{sec:author-contributions}
Yilin Zhai was responsible for the overall conceptualization, data analysis, experimental design, and drafting of the manuscript; Hongyuan Shi participated in research discussions, provided valuable guidance and revisions; Zaijin You assisted with data discussions and contributed to several sections of the manuscript.  All authors have read and approved the final version of the manuscript.

\section{Conflict of Interest}
\label{sec:conflict-of-interest}
The authors declare that the research was conducted in the absence of any commercial or financial relationships that could be construed as a potential conflict of interest.

\section{Data Availability}
\label{sec:data}
The NDBC buoy records analysed in this study are publicly available from the National Data Buoy Center (\url{https://www.ndbc.noaa.gov}). The full hyperparameter search results for all 90 (family $\times$ context length) best-trial configurations in both the single-buoy and multi-buoy phases are provided as Supplementary Material ({\tt trials.csv}).

\bibliographystyle{elsarticle-harv}

\bibliography{refs.bib}
\end{document}